\documentclass{WileyMSP-template}

\usepackage{placeins}

\begin{document}

\pagestyle{fancy}

\title{Unraveling Nanostructured Spin Textures in Bulk Magnets}

\maketitle


\author{Philipp Bender*}
\author{Jonathan Leliaert}
\author{Mathias Bersweiler}
\author{Dirk Honecker}
\author{Andreas Michels}


\dedication{}


\begin{affiliations}
Dr. P. Bender, Dr. M. Bersweiler, Dr. D. Honecker, Prof. A. Michels\\
Department of Physics and Materials Science, University of Luxembourg, 162A~avenue de la Fa\"iencerie, L-1511 Luxembourg, Grand Duchy of Luxembourg\\
Email Address: philipp.bender@uni.lu

Dr. J. Leliaert\\
Department of Solid State Sciences, Ghent University, Krijgslaan 281/S1, 9000 Ghent, Belgium

\end{affiliations}



\begin{abstract}

One of the key challenges in magnetism remains the determination of the nanoscopic magnetization profile within the volume of thick samples, such as permanent ferromagnets. Thanks to the large penetration depth of neutrons, magnetic small-angle neutron scattering (SANS) is a powerful technique to characterize bulk samples. The major challenge regarding magnetic SANS is accessing the real-space magnetization vector field from the reciprocal scattering data. In this letter, a fast iterative algorithm is introduced that allows one to extract the underlying two-dimensional magnetic correlation functions from the scattering patterns. This approach is used here to analyze the magnetic microstructure of Nanoperm, a nanocrystalline alloy which is widely used in power electronics due to its extraordinary soft magnetic properties. It can be shown that the computed correlation functions clearly reflect the projection of the three-dimensional magnetization vector field onto the detector plane, which demonstrates that the used methodology can be applied to probe directly spin-textures within bulk samples with nanometer-resolution.

\end{abstract}

\

Nanostructured magnetic materials attract much interest thanks to the unique magnetic properties that can arise when the structural units (e.g. particles, crystallites, or film layers) are reduced below a characteristic intrinsic magnetic length scale of the system \cite{fischer2020launching}.
The prototype of spatially-localized magnetic objects are magnetic nanoparticles, which have a single domain magnetization below a material-specific size of typically a few tens of nanometers \cite{skomski2003nanomagnetics}.
For larger sizes or deviation from spherical shape, even defect-free magnetic nanoparticles can display more complex spin structures such as curling-, flower-, or vortex-states \cite{pinilla2016switching}.
In general, to fully understand the complex interplay between the structural and magnetic properties of nanostructured magnetic materials, the determination of the internal magnetization profile remains a key challenge \cite{fernandez2017three}.
In case of individual, free-standing magnetic structures, such as nanoparticles, or micrometer-sized pillars and discs, the internal magnetization profile can be determined, e.g. by advanced electron microscopy \cite{phatak2014visualization,tanigaki2015three,gatel2015size} or X-ray scattering techniques \cite{streubel2015retrieving,donnelly2017three,donnelly2020time}.
In particular, X-ray nanotomography enables nowadays the reconstruction of the three-dimensional (3D) magnetization vector field within magnetic microstructures such as shape-memory elastomers \cite{testa2019magnetically}.
Furthermore, electrons and X-rays can be used to investigate the interparticle moment coupling in planar 2D assemblies of interacting magnetic nanoparticles \cite{varon2013dipolar,chesnel2018unraveling}.
However, for thick, mm-sized 3D samples neither electrons nor X-rays are normally suitable to access buried magnetic structures due to their small penetration depths, and thus the internal magnetization profile of most bulk magnets or 3D nanoparticle assemblies is not resolvable with these techniques.
To reveal the complex magnetization profile within 3D magnetic systems small-angle neutron scattering (SANS) can be employed \cite{muhlbauer2019magnetic}.
The general advantage of neutrons as an important probe of magnetism is their large penetration depths, which allows the characterization of bulk materials with thicknesses of up to several millimeters \cite{kardjilov2008three,manke2010three}.

In Nanoperm, a technologically-relevant nanostructured magnetic alloy \cite{suzuki06}, small Fe nanocrystallites are embedded in a soft magnetic, amorphous matrix.
This system provides a complex testing ground because its magnetization consists of approximately single-domain Fe particles, surrounded by a magnetically softer matrix that is distorted due to the dipolar stray fields \cite{vecchini2005neutron}.
In previous SANS studies of Nanoperm it was shown that the magnetodipolar stray fields of the Fe nanocrystals cause characteristic anisotropies in the magnetic SANS patterns \cite{michels2005dipole,michels2006dipolar}.
However, in general, the key challenge regarding magnetic SANS remains accessing the real-space magnetization vector field from the reciprocal scattering data.
In most studies, data analysis is done by analyzing 1D sectors or radial averages, e.g by fitting the data to a particular model in reciprocal space \cite{krycka2010core,disch2012quantitative}, or by determining model-independently the real-space 1D correlation functions \cite{bersweiler2019size,bender2019supraferromagnetic}.
But due to the anisotropic nature of magnetic scattering, reducing the analysis to 1D essentially means a loss of information.
Moreover, in many studies structural form-factor models, adapted from nuclear SANS, are utilized, which fail to account for the existing spin inhomogeneity inside magnetic nanostructured systems.
Only recently the analysis of the total (magnetic and/or nuclear) 2D patterns was introduced, either by directly calculating the cross section in reciprocal space \cite{alves2017calculation,zakutna2019morphological} or by determining the real-space 2D correlation functions \cite{mettus2015small,fritz2015interpretation}.

In this paper, we introduce a new method to extract the underlying 2D correlation functions from 2D magnetic SANS patterns.
This approach can be readily applied for the model-free analysis of diffuse magnetic SANS data in various research fields, including e.g. multiferroic alloys \cite{bhatti2012small}, permanent magnets \cite{perigo2015magnetic}, multilayer systems \cite{dufour2011nanometer}, magnetic steels \cite{bischof2007influence}, nanogranular magnetic films \cite{alba2016magnetic}, nanowire arrays \cite{gunther2014magnetic,grutter2017complex}, ferrofluids \cite{wiedenmann2011relaxation}, and magnetic nanoparticles \cite{ito2007effect,dennis2015internal}.
Furthermore, the numerical algorithm for the extraction of the 2D magnetic correlation function can be easily transferred to other experimental techniques where correlation functions are measured, like spin-echo neutron scattering to study dynamics in the neV-energy range \cite{strobl2012tof,franz2019longitudinal}, or pair distribution function analysis in diffraction for information on atomic disorder \cite{frandsen2014magnetic}, and other complementary X-ray techniques, such as resonant soft X-ray magnetic scattering \cite{scherz2007phase} and X-ray photon correlation spectroscopy \cite{seaberg2017nanosecond}.
In this study, our approach is used to analyze the magnetic SANS data of Nanoperm and we show that the derived correlation functions nicely reflect the field-dependent, real-space, nanoscale magnetization configuration within the bulk samples predicted by micromagnetic simulation and theory.

We obtained the field-dependent and purely magnetic SANS cross sections $I_m(\mathbf{q})$ from the experimental data by subtracting the total (nuclear and magnetic) SANS cross section measured at saturation ($\mu_0H\cong2$\,T) from the measurements at intermediate field strengths \cite{michels2006dipolar}:
\begin{equation}\label{eq1}
I_m(\mathbf{q})\propto |\widetilde{M}_x|^2+|\widetilde{M}_y|^2\mathrm{cos}^2\Theta-\left(\widetilde{M}_y\widetilde{M}^*_z+\widetilde{M}^*_y\widetilde{M}_z\right)\mathrm{sin}\Theta\mathrm{cos}\Theta.
\end{equation} 
The scattering vector $\mathbf{q}$ is defined in the detector $yz$-plane.
In writing down Equation\,\ref{eq1} it is assumed that the sample is in the approach-to-saturation regime, and that $M(2\,\mathrm{T})\approx M_S$, with $M_S$ being the saturation magnetization; $\Theta$ is the angle between $\mathbf{q}$ and $\mathbf{H}$, and $\widetilde{M}_x(\mathbf{q})$, $\widetilde{M}_y(\mathbf{q})$, $\widetilde{M}_z(\mathbf{q})$ are the Fourier transforms of the magnetization components $M_x(\mathbf{r})$, $M_y(\mathbf{r})$, $M_z(\mathbf{r})$ of the real-space magnetization vector field, where the asterisk $'*'$ indicates the complex-conjugate.
Note that the Fourier components $\widetilde{M}_{x,y,z}(\mathbf{q})$ can be anisotropic, which severely complicates a decoupling of the individual scattering contributions in Equation\,\ref{eq1}.

In principle, the real-space 2D magnetic correlation function $P(\mathbf{r})=rC(\mathbf{r})$, with $C(\mathbf{r})$ being the autocorrelation function in case of nuclear scattering, can be extracted from the experimental reciprocal scattering data $I_m(\mathbf{q})$ \textit{via} a direct Fourier transform \cite{mettus2015small}.
For the analysis of nuclear scattering patterns, however, usually indirect approaches are applied where the inverse problem is solved \cite{fritz2013two,fritz2015interpretation,bender2019using}, and which can be readily adapted to magnetic SANS. 
The challenge is to extract good and robust estimations for $P(\mathbf{r})$ from the noisy data, also in case of restricted $q$-ranges as is usually the case in experiment.
An additional problem regarding the evaluation of 2D scattering patterns is the necessary computation time related to processing the large matrices involved (i.e. the data and the 2D correlation function).
Here, we introduce an iterative method (called Kaczmarz' algorithm \cite{KAC-37}) to solve this ill-conditioned problem, which was already used successfully for the fast analysis of magnetic particle imaging, magnetometry and magnetorelaxometry data of magnetic nanoparticle ensembles \cite{schmidt2017finding,leliaert2017interpreting}.

For $\mathbf{k}||\mathbf{e_x}$, the 2D scattering intensity can be written in polar coordinates as $I_m(q_y,q_z)=I_m(q,\Theta)$, with $q=|\mathbf{q}|$ and $\Theta=\mathrm{arctan}(q_y/q_z)$.
The 2D scattering pattern has $N$ pixels, and for each pixel '$i$' (i.e. data point) it can be expressed as:
\begin{equation}\label{eq2}
I(q_i,\Theta_i)=\sum_{j=1}^K A_{ij}P(r_j,\varphi_j).
\end{equation}
The angle $\varphi$ specifies the orientation of $\mathbf{r}$ in the $yz$-plane, and the extracted 2D distribution function $P(\mathbf{r})$ is given by $P(r,\varphi)=C(r,\varphi)r$ \cite{fritz2013two}.
The matrix $\mathbf{A}$ in Equation\,\ref{eq2} is the data transfer matrix, which, in case of the 2D indirect Fourier transform, has the elements \cite{mettus2015small,fritz2013two,fritz2015interpretation,bender2019using} 
\begin{equation}\label{eq3}
A_{ij}=\mathrm{cos}\left(q_ir_j\mathrm{cos}\left(\Theta_i-\varphi_j\right)\right)\Delta r_j\Delta\varphi_j.
\end{equation}
As is typical in such an analysis, we use a linear spacing for the pre-determined $r$- and $\varphi$-vectors.
We use the following algorithm from Kaczmarz to update the elements $P(r_j,\varphi_j)$ after each iteration according to:
\begin{equation}\label{eq4}
P^{k+1}(r_j,\varphi_j)=P^{k}(r_j,\varphi_j)+\frac{I(q_i,\Theta_i)-\left(A_i\cdot P^{k}(r_j,\varphi_j)\right)}{\sigma||A_i||^2}\overline{A_i},
\end{equation}
where $A_i$ is the $i$th row of the matrix $\mathbf{A}$, $\overline{A_i}$ is its transpose, $k$ is the iteration number, and one iteration contains a sweep over all rows $i$.
Hereby, we shuffle randomly through all rows $A_i$, and normalize the residuals (i.e. $I(q_i,\Theta_i)-\left(A_i\cdot P^{k}(r_j,\varphi_j)\right)$) to $\sigma=\sqrt{I(q_i,\Theta_i)}$, similar to a weighted least-squares fit. 
The Kaczmarz algorithm can be of course also used to determine the 1D correlation functions from 1D data sets (e.g. the radial average $I(q)=1/(2\pi)\int_0^{2\pi}I(q,\Theta)\mathrm{d\Theta}$ or individual sectors).
In this case, Equation\,\ref{eq4} is applied to determine $P^{k+1}(r_j)$ with $A_{ij}=\mathrm{sin}\left(q_ir_j\right)/\left(q_ir_j\right)\Delta r_j$ being the matrix elements \cite{bender2017structural}.

\begin{figure}[t]
\includegraphics[width=1\columnwidth]{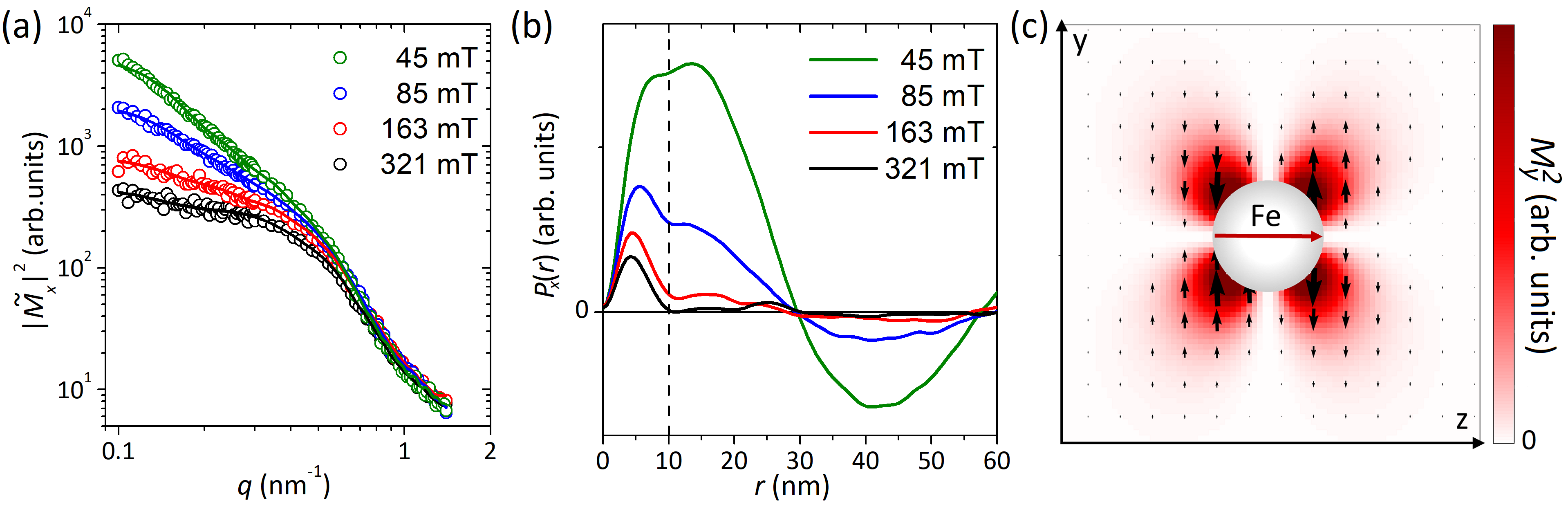}
\caption{\label{Fig1}
(a) Experimental data for the 1D cross section $|\widetilde{M}_x|^2$ (i.e. the vertical sector of the 2D scattering patterns, $\Theta=90^{\circ}\pm10^{\circ}$) for different field strengths, plotted over the  accessible $q$-range ($0.1-1.2\,\mathrm{nm^{-1}}$). The solid lines are the fits based on Kazcmarz' algorithm.
(b) The 1D correlation functions $P_x(r)$ ($r=0-100$\,nm in 1-nm steps) extracted from $|\widetilde{M}_x|^2$ \textit{via} Kaczmarz' algorithm  after $k=50$ iteration steps.
(c) Simulated real-space magnetization profile of the component $M_y^2$ around a spherical iron particle embedded within a soft magnetic matrix. The black arrows indicate the orientation of $M_y$, which follows $M_y \propto\Delta M\mathrm{sin}\varphi\mathrm{cos}\varphi/r^3$. The field ($\mu_0H=163\,\mathrm{mT}$) was applied along the $z$-direction.}
\end{figure}

Before focusing on the 2D data, we will first use this approach to analyze the 1D cross section $|\widetilde{M}_x|^2$.
The transversal magnetization $|\widetilde{M}_x|^2$ is of interest because it can be easily extracted from the vertical sectors of the 2D scattering patterns ($\Theta=90^{\circ}\pm 10^{\circ}$, Equation\,\ref{eq1}). 
In \textbf{Figure\,\ref{Fig1}}(a) we show the field dependence of $|\widetilde{M}_x|^2$, whereas Figure\,\ref{Fig1}(b) displays the 1D correlation functions $P_x(r)$.
As can be seen, at the highest field strength $P_x(r)$ exhibits a well pronounced peak for $0<r<10\,\mathrm{nm}$.
It is safe to assume that this peak corresponds to the individual Fe crystallites, which have a size of around 12\,nm and are in a single-domain state.
This peak indicates that at 321\,mT the magnetization inside the crystallites slightly deviates from perfect alignment along the field direction, probably due to the local magnetocrystalline anisotropy.
With further decreasing field strength the magnitude of the peak increases but its position remains the same indicating a further tilting of the particle moments along the easy axis.
Even at the lowest field we still see a shoulder at around 5\,nm which is attributed to the single-domain Fe crystallites.
In addition to the increase in peak intensity, at decreasing field strength we also observe progressively more deviation of $P_x(r)$ from zero for $r>10\,\mathrm{nm}$. 
This corresponds to the increased slope we observe for $|\widetilde{M}_x|^2$ in the low $q$-range in Figure\,\ref{Fig1}(a), and indicates the formation of an inhomogeneous magnetization profile around the crystallites.
With decreasing external field strength the perturbations of the magnetization increases within the vicinity of the Fe crystallites.

To verify the strong influence of the stray field on the local magnetization configuration we simulated the magnetic nanostructure of Nanoperm with MuMax3 \cite{vansteenkiste2014design}.
Figure\,\ref{Fig1}(c) shows the squared $y$-component of the magnetization vector field at 163\,mT (the black arrows indicate the orientation of $M_y$).
The stray field of the Fe sphere with the approximate functional form \cite{vecchini2005neutron}
\begin{equation}\label{eq5}
M_y \propto\Delta M\mathrm{sin}\varphi\mathrm{cos}\varphi/r^3.
\end{equation} 
results in a perturbation of the magnetization of the surrounding matrix picking up the symmetry of the stray field.
In Equation\,\ref{eq5}, $\Delta M$ denotes the jump in the magnitude of the magnetization at the particle-matrix interface, which is about 1.5\,T for Nanoperm.
It is important to note that the expected behavior for $M_y$ and $M_x$ is identical since the symmetry is broken along the field direction, i.e. the $z$-direction.
With reference to Equation\,\ref{eq1} the predicted effect of the dipole fields on the Fourier transform of the magnetization is that $\widetilde{M}_y\approx\hat{M}_y \mathrm{sin}\Theta\mathrm{cos}\Theta$, where $\hat{M}_y$ is the angular independent (i.e. isotropic) amplitude of $\widetilde{M}_y$.
Moreover, we can assume that for a statistically-isotropic microstructure also $\widetilde{M}_z$ is angular independent (i.e. $\widetilde{M}_z=\hat{M}_z$) and therefore we can write for the cross-term in Equation\,\ref{eq1} at first approximation $\hat{M}_z\hat{M}_y\mathrm{sin}^2\Theta\mathrm{cos}^2\Theta$.

\begin{figure}[t]
\includegraphics[width=1\columnwidth]{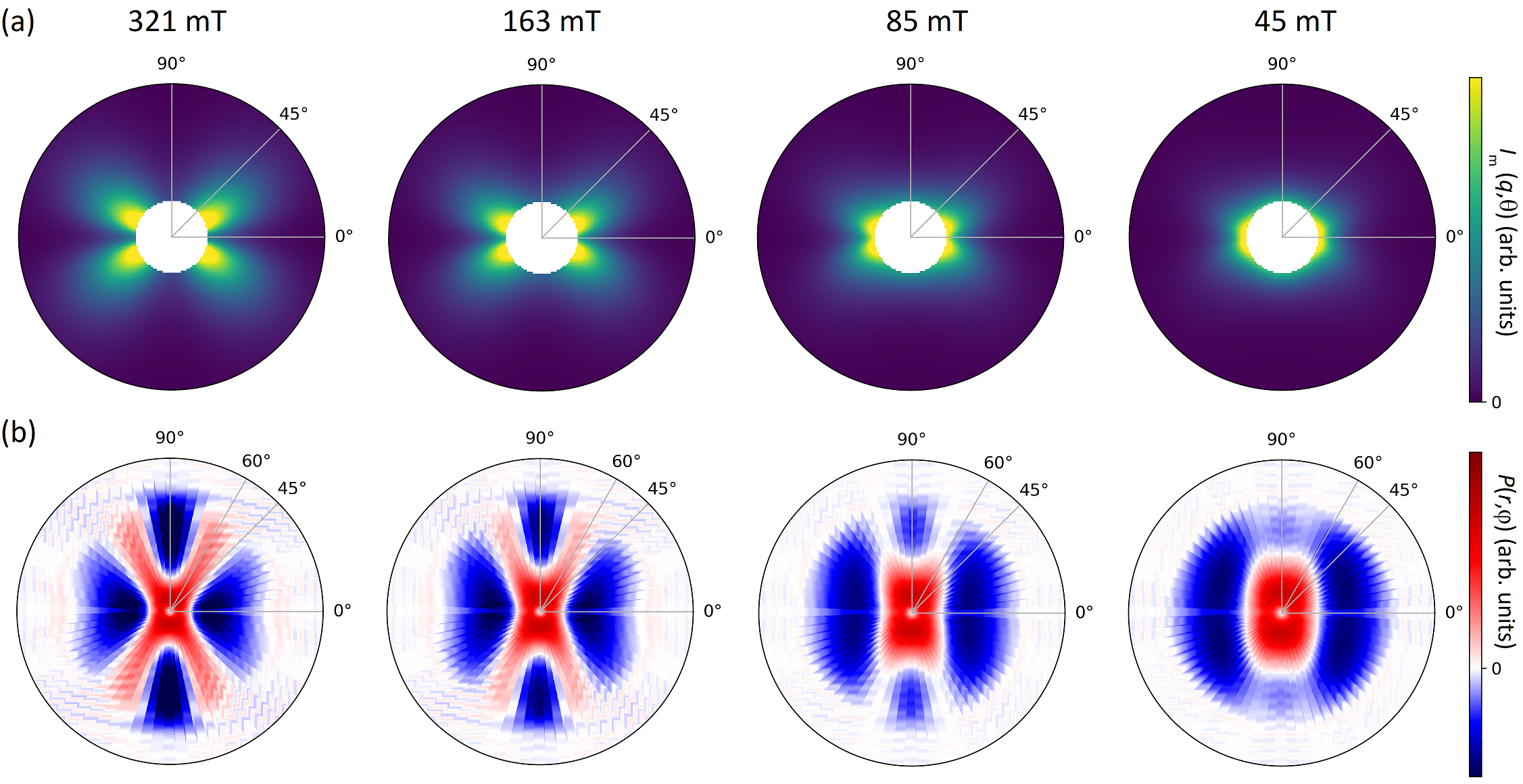}
\caption{\label{Fig2}
(a) Theoretical 2D SANS patterns $I_m(\mathbf{q})$ calculated for the four field strengths $\mu_0H=321$, 163, 85, 45\,mT, using the same model and material parameters as in \cite{honecker2013analysis}.
The Fe crystallites are assumed to be well-separated spheres and lognormally distributed with a mean size of 10\,nm and $\sigma=0.2$.
The plotted $q$-range is $0.1-0.4\,\mathrm{nm^{-1}}$, but the total $q$-range we used for the calculations was $0.1-1.2\,\mathrm{nm^{-1}}$.
(b) Corresponding 2D correlation functions $P(\mathbf{r})$ extracted from the above scattering patterns $I_m(\mathbf{q})$ using the Kaczmarz algorithm ($k=50$ iterations, see Equation\,\ref{eq4}).
The patterns $P(r,\varphi)$ were determined for $r=0-100$\,nm in 1-nm steps and $\varphi=0-360^\circ$ in $5^\circ$-steps. The plotted $r$-range is $0-80$\,nm.
}
\end{figure}

To describe the scenario depicted in Figure\,\ref{Fig1}(c), an analytical theory for $I_m(\mathbf{q})$ was developed in \cite{honecker2013analysis,honecker2013theory}, which is valid in the approach to saturation (see Equation\,\ref{eq1}).
\textbf{Figure\,\ref{Fig2}}(a) displays the calculated 2D patterns using this model for the field strengths of 321, 163, 85 and 45 mT, and in Figure\,\ref{Fig2}(b) we plot the corresponding 2D correlation functions which we extracted from the synthetic scattering data using the Kaczmarz algorithm (see Equation\,\ref{eq4}).
It can be seen that at 321\,mT the scattering pattern is dominated by the $\mathrm{sin}^2\Theta\mathrm{cos}^2\Theta$-term.
With decreasing field strength the signature of the $\mathrm{sin}^2\Theta\mathrm{cos}^2\Theta$ term vanishes which indicates an increasing contribution by the $|\widetilde{M}_x|^2$ and $|\widetilde{M}_y|^2$ terms (Equation\,\ref{eq1}).
Consequently, the corresponding 2D correlation functions vary significantly with field strength.
Although $P(\mathbf{r})$ is not directly the autocorrelation function of the real-space magnetization vector field \cite{mettus2015small}, the extracted correlation functions clearly reflect the characteristic features of the real-pace 3D magnetization profile. 
At high field strengths, i.e. 321 and 163\,mT, $P(\mathbf{r})$ displays a pronounced anisotropy with maxima along $\Theta=60^\circ$, which follows the dipolar stray field of the Fe crystallites (see also Figure\,\ref{Fig1}(c)).
With decreasing field strength, the anisotropy of $P(\mathbf{r})$ changes and is elongated along the vertical direction ($\Theta=90^\circ$) in agreement with theory \cite{mettus2015small}.

\begin{figure}[t]
\includegraphics[width=1\columnwidth]{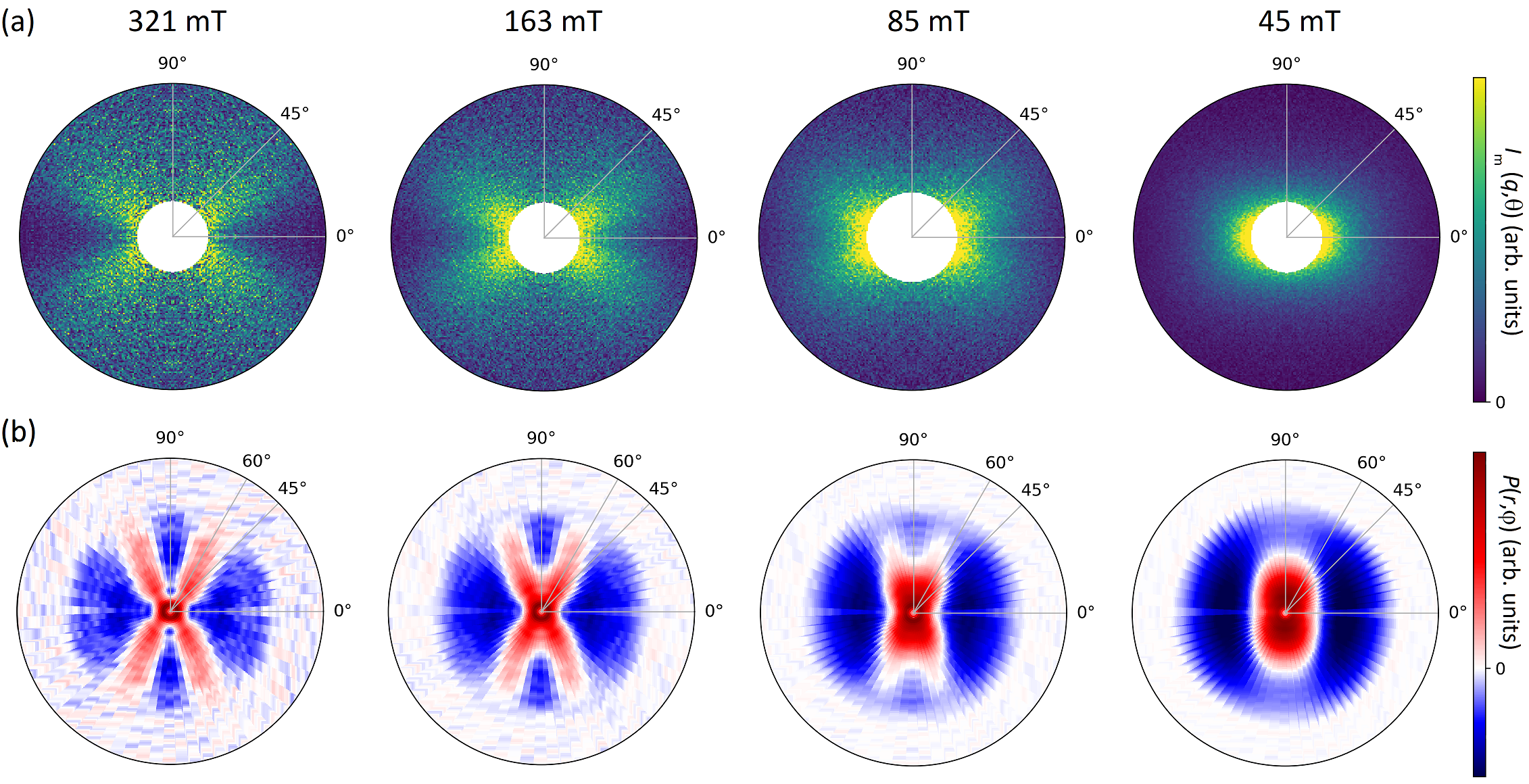}
\caption{\label{Fig3}
(a) Experimental 2D SANS patterns $I_m(\mathbf{q})$ measured at the four field strengths $\mu_0H=321$, 163, 85, 45\,mT ($q=0.1-0.4\,\mathrm{nm^{-1}}$). 
(b) Corresponding 2D correlation functions $P(\mathbf{r})$ extracted from the above scattering patterns $I_m(\mathbf{q})$ using  the Kaczmarz algorithm ($k=50$ iterations, see Equation\,\ref{eq4}).
The patterns $P(r,\varphi)$ were determined for $r=0-100$\,nm in 1-nm steps and $\varphi=0-360^\circ$ in $5^\circ$-steps. The plotted $r$-range is $0-80$\,nm.}
\end{figure}

In \textbf{Figure\,\ref{Fig3}} we show the experimental 2D scattering data for all four field strengths, as well as the corresponding 2D correlation functions.
It is evident that the experimentally observed angular anisotropies are in excellent agreement with the theoretical predictions in Figure\,\ref{Fig2}.
For all four field strengths we obtain basically the identical 2D correlation function.
This demonstrates that the Kaczmarz algorithm can be employed to robustly extract the underlying 2D magnetic correlation functions from noisy experimental magnetic SANS data.

To summarize, we have introduced a procedure for the analysis of magnetic SANS data to probe the magnetization configuration of bulk samples.
We applied this approach to characterize the magnetic microstructure of the nanocrystalline ferromagnet Nanoperm.
By subtracting the nuclear and magnetic scattering at saturation from the SANS data at intermediate magnetic field strengths, we obtained the purely magnetic SANS intensities $I_m(\mathbf{q})$.
We employed Kaczmarz' algorithm to extract the 2D magnetic correlation functions from the scattering patterns $I_m(\mathbf{q})$.
By comparing our results with micromagnetic simulations and theoretical calculations we can show that the extracted correlation functions accurately reflect the real-space magnetization distribution following the dipolar stray fields around Fe nanocrystallites.

This study highlights that the 2D correlation functions derived from magnetic SANS data carry useful and important information regarding nanostructured spin textures within bulk magnets.
As clarified before, the correlation function does not directly coincide with the auto-correlation function of the magnetization vector field, but it is related to the projection of the 3D real-space magnetization configuration onto the 2D detector plane.
Its determination is a unique way to access the internal magnetization profile buried within nanostructured bulk magnets.
As discussed in the introduction, the large penetration depth of neutrons enables probing sample volumes in the $\mathrm{mm^3}$-range, in contrast to electron and X-ray techniques that are localized, surface-sensitive probes and which can be only used for thin samples or free-standing structures.
Diffuse magnetic SANS is not sensitive to the crystalline, atomic structure of the sample, but measures with nanometer-resolution the spatially and time-averaged magnetic microstructure over the mesoscale ($\sim$ 1 - 500\,nm) regime, which is often the key to realize specific properties and functions of structured materials.
We envision that our approach becomes a powerful tool for the model-free analysis of diffuse magnetic SANS data in various research fields, 
and we believe that the presented approach to extract 2D magnetic correlation functions can be easily transferred to other experimental techniques where magnetic correlation or pair distance distribution functions are measured, like e.g. spin-echo neutron scattering.

\section*{Experimental Section}

\textit{Sample Preparation:}
The Nanoperm ($\mathrm{Fe_{89}Zr_7B_3Cu}$) sample was prepared by melt spinning and subsequent annealing for 1\,h at 745\,K.
The sample had Fe crystallites with an average size of 12\,nm according to X-ray diffraction and electron microscopy.
Several ribbons with a thickness of 0.02\,mm were stacked on top of each other and mounted on the sample holder for the SANS experiment

\textit{SANS measurements:}
The magnetic-field-dependent, unpolarized SANS measurements were performed at room temperature on the SANS-2 instrument at GKSS, Geesthacht, Germany, using an average neutron wavelength of $\lambda = 0.58$\,nm with a wavelength spread of 10\%. 
Each measurement took several minutes to reach sufficient statistics.
The neutron beam had a diameter of 8\,mm and the total sample thickness was $\sim$ 0.2\,mm, thus the probed sample volume was in the $\mathrm{mm^3}$-range.
In this study we focus our analysis on the SANS measurements within a $q$-range of $q=0.1-1.2\,\mathrm{nm^{-1}}$.
The homogeneous magnetic field $\mathbf{H}||\mathbf{e}_z$ was applied normal to the incident neutron beam $\mathbf{k}||\mathbf{e}_x$ and in the plane of the sample.
The measurements were performed at field strengths of $\mu_0H=321$, 163, 85, and 45\,mT (all in the approach to saturation), as well as in the saturated state (at $\mu_0H\cong2$\,T), which serves as the background signal.

\textit{Micromagnetic simulation:}
We simulated the magnetic nanostructure of Nanoperm with MuMax3 \cite{vansteenkiste2014design}, using the materials parameters for Nanoperm from \cite{honecker2013analysis}.

\medskip
\textbf{Acknowledgements} \par 
P. B. acknowledges financial support from the National Research Fund of Luxembourg (CORE SANS4NCC grant). 
J. L. is supported by the Research Foundation – Flanders (FWO) through a postdoctoral fellowship. 
We thank Kiyonori Suzuki (Monash University) for providing the Nanoperm sample and Klaus Pranzas (Helmholtz-Zentrum Geesthacht) for assisting in the SANS experiment.

\medskip

%
\bibliographystyle{MSP}
\bibliography{PBender_Bib}


\FloatBarrier
\begin{figure}[b]
\textbf{Table of Contents}\\
\medskip
  \includegraphics{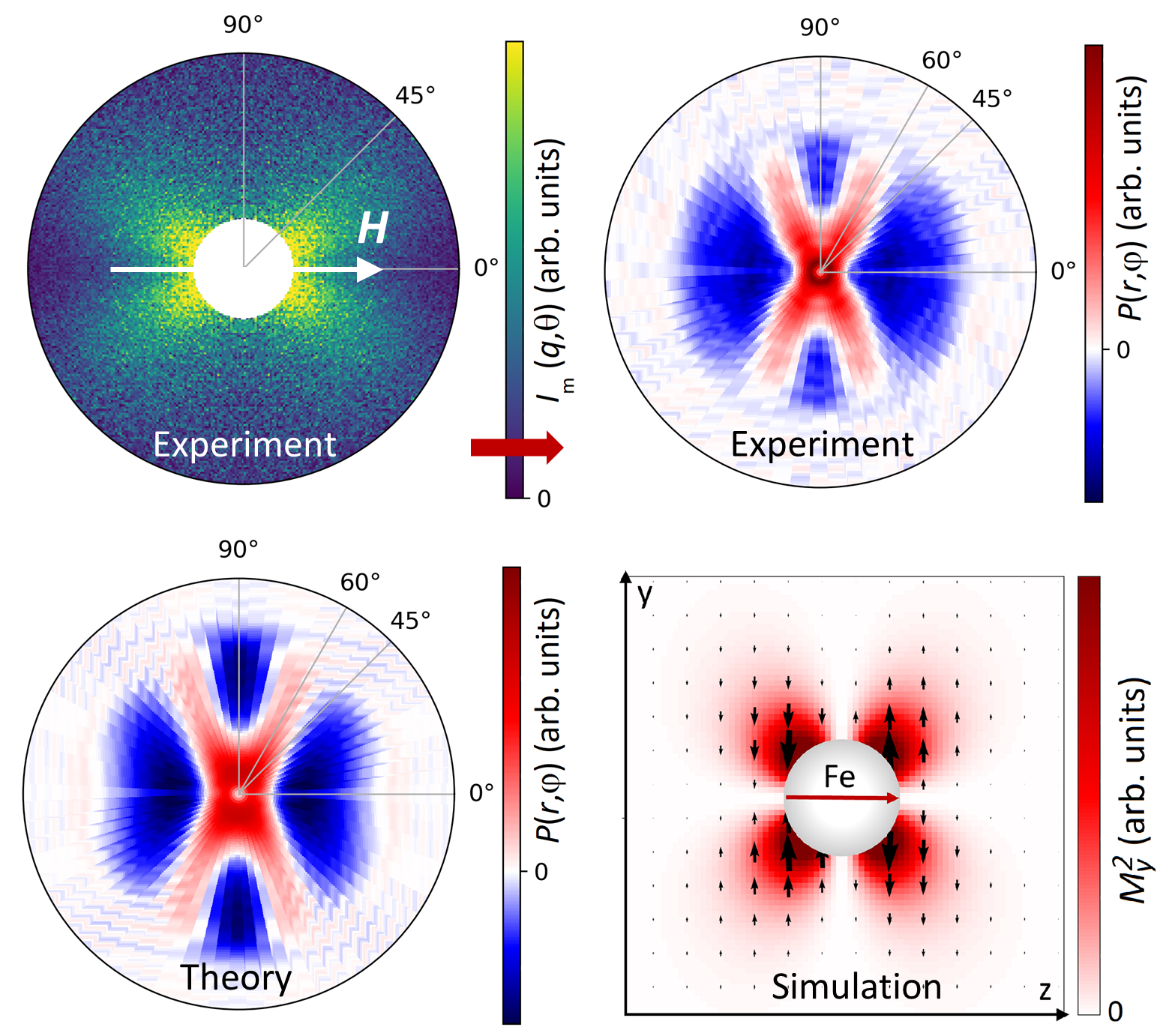}
  \medskip
  \caption*{A fast iterative algorithm is introduced to extract the underlying two-dimensional magnetic correlation functions from magnetic SANS patterns. This approach is used to probe the nanoscopic spin textures within the nanostructured soft magnetic alloy Nanoperm. By comparing the results with micromagnetic simulations and theoretical calculations it can be verified that the extracted correlation functions reflect the real-space magnetization distribution.}
\end{figure}

\end{document}